\documentclass[twocolumn,showpacs,prb]{revtex4}
\usepackage{graphicx}
\usepackage{dcolumn}
\usepackage{bm}

\begin{document}
\title{Low-Temperature Permittivity of Insulating Perovskite Manganites}
\author{J. L. Cohn,$^1$ M. Peterca,$^{1,\dagger}$, and J. J. Neumeier$^2$}
\affiliation{$^1$ Department of Physics, University of Miami, Coral
Gables, Florida 33124}
\affiliation{$^2$ Department of Physics, Montana State University,
Bozeman, Montana}

\begin{abstract}
Measurements of the low-frequency ($f\leq 100$ kHz) permittivity ($\varepsilon$)
and conductivity ($\sigma$) at $T\lesssim 150 K$
are reported for La$_{1-x}$Ca$_x$MnO$_3$ ($0\leq x\leq 1$) and
Ca$_{1-y}$Sr$_y$MnO$_3$ ($0\leq y\leq 0.75$) having antiferromagnetic,
insulating ground states covering a broad range of Mn valencies
from Mn$^{3+}$ to Mn$^{4+}$. Static dielectric constants are
determined from the low-$T$ limiting behavior.  With increasing $T$,
relaxation peaks associated with charge-carrier
hopping are observed in the real part of the permittivities and
analyzed to determine dopant binding energies. The data are
consistent with a simple model of hydrogenic impurity levels
and imply effective masses $m^{\ast}/m_e\sim 3$ for the
Mn$^{4+}$ compounds.  Particularly interesting is a large dielectric constant
($\varepsilon_0\sim 100$) associated with the C-type antiferromagnetic
state near the composition La$_{0.2}$Ca$_{0.8}$MnO$_3$.

\end{abstract}
\pacs{75.47.Lx, 75.50.Ee, 77.22.Ch,77.22.Gm, 71.55.-i}
\maketitle

\section{\label{sec:Intro} Introduction}

Recently, the lightly electron-doped manganites, e.g.
Ca$_{1-x}$L$_x$MnO$_3$ ($L$ is a Lanthanide), have been shown to exhibit a novel phase
separated ground state, composed of distinct crystallographic and
magnetic phases on a mesoscopic scale.\cite{NeumeierCohn,RaveauGroup,Santhosh,
Aliaga,LingGranado,Machida}  Detailed neutron diffraction studies\cite{LingGranado}
of Ca$_{1-x}$La$_x$MnO$_3$ ($x\leq 0.2$) indicate that the heterogeneity
of this system is intrinsic, associated with an extremely fine balance between competing
ferromagnetic (FM) double-exchange and antiferromagnetic (AF) superexchange interactions.

This paper reports investigations of the compositional dependence of the static
dielectric constant ($\varepsilon_0$) in the Mn$^{4+}$-rich
portion of the manganite phase diagram, accessible through low-frequency ($f\leq 100$~kHz)
impedance measurements at low temperature ($T\geq 2$~K).
Very few studies of the permittivity of manganites have been reported.\cite{LunkenheimerLSMO,LunkenheimerPSCMO}
Specimens for which homogeneous, insulating ground states predominate
are the particular focus: The A-type AF phase (LaMnO$_3$),
the Wigner-crystal AF phase (La$_{1/3}$Ca$_{2/3}$MnO$_3$), the C-type AF phase (La$_{0.2}$Ca$_{0.8}$MnO$_3$),
and the G-type AF phase (Ca$_{1-y}$Sr$_y$MnO$_3$).
In general, $\varepsilon_0$ is an important parameter for models of phase
separation involving the segregation of doped charge carriers on a
mesoscopic scale, relevant for some compositions near to those investigated here.
It is also a key parameter in determining polaronic binding
energies.\cite{Alexandrov} In addition, impedance measurements provide
direct information about the charge carriers since carrier hopping
yields a dipolar contribution to the permittivity.

\section{\label{sec:Expt} Experiment}

Polycrystalline La$_{1-x}$Ca$_x$MnO$_3$ (LCMO) and
Ca$_{1-y}$Sr$_y$MnO$_3$ (CSMO) specimens were prepared by standard
solid-state reaction; the preparation methods and magnetization
and transport measurements are reported
elsewhere.\cite{NeumeierCohn,CohnNeumeier} Powder x-ray
diffraction revealed no secondary phases and iodometric titration,
to measure the average Mn valence, indicates the oxygen content of
all specimens falls within the range 3.00$\pm$0.01.
\par
AC impedance measurements were conducted with an HP4263B LCR meter
at frequencies $f=$~100~Hz, 120~Hz, 10~kHz, 20~kHz, 100~kHz using
a 4-terminal pair arrangement. Reliable measurements of
$\varepsilon$ were restricted to $T\lesssim 160$ K where the
capacitive reactance was sufficiently large ($\gtrsim 0.1\Omega$).
Typical specimen dimensions were 3$\times$1.0$\times$0.5 mm.
Silver paint electrodes were applied on the largest, polished
faces of the specimens and annealed at 300\,$^{\circ}$C for 2 h to
improve contact resistance.  Contact capacitance can lead to
apparently large values\cite{Lunkenheimer} of $\varepsilon$ and
thus some care is required to distinguish the true response of the
sample. To rule out the influence of contacts, the impedances of
several specimens were remeasured after further polishing to
reduce the electrode spacing by at least a factor of two; in all
cases the low-temperature data agreed within geometric
uncertainties of $\pm 10$\%. The results were also independent of
applied DC bias from 0V-2V, and ac voltage in the range 50mV-1V.

\section{\label{sec:ResultsDisc} Results and Analysis}
\subsection{\label{Equations} Relations}

\begin{figure}
\vskip -.1in
\includegraphics[width = 3.5in]{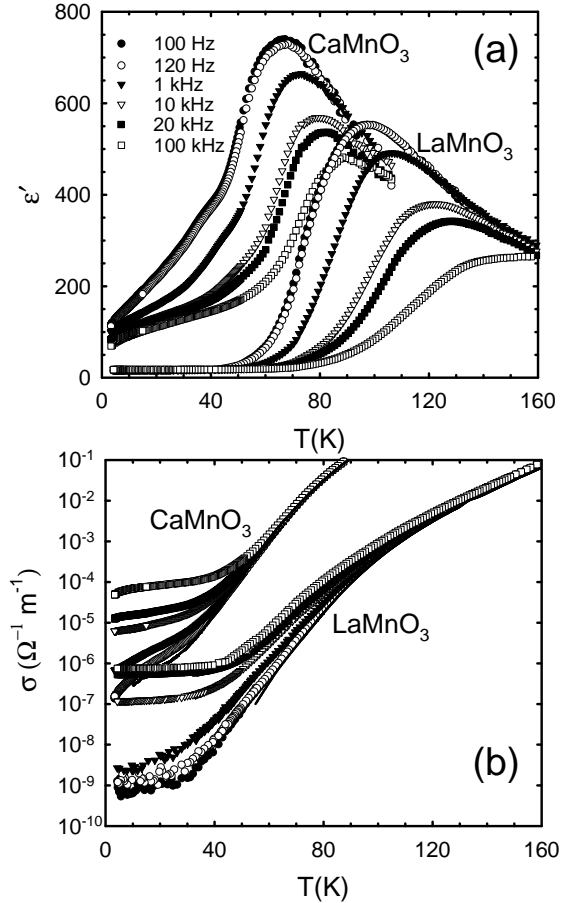}%
\vskip -.2in
\caption{(a) $\varepsilon^{\prime}$ and (b) $\sigma$ {\it vs.}
temperature for CaMnO$_3$ and LaMnO$_3$.  Solid curves in (b) are
dc conductivities.} \label{LMOCMOepsdata}
\end{figure}

Quite generally, the complex dielectric permittivity of a solid,
$\varepsilon=\varepsilon^{\prime}-i\varepsilon^{\prime\prime}$,
can be expressed as,
$\varepsilon=\varepsilon_{\infty}+\varepsilon_l+\varepsilon_d$.
$\varepsilon_{\infty}$ is the high-frequency dielectric constant
associated with displacements of ionic charge distributions
relative to their nuclei.  The lattice contribution,
$\varepsilon_l$, arises from displacements of ions and their
charge distributions. $\varepsilon_d$ represents a dipolar
contribution, associated in the present materials with
charge-carrier hopping.  $\varepsilon_{\infty}$ and
$\varepsilon_l$ are generally frequency- and temperature-independent at low $T$.
The frequency-dependent dipolar conductivity is described by a power
law,\cite{PowerlawSigma,Jonscher} and is reflected in the
dielectric loss ($\varepsilon_d^{\prime\prime}$),
\begin{eqnarray}
\sigma_d(\omega)=\sigma_0\omega^s=\omega\epsilon_0\varepsilon_d^{\prime\prime}(\omega),
\label{SigmaofOmega}
\end{eqnarray}
where $\omega (=2\pi f)$ is the angular frequency, $\sigma_0$ is
generally weakly $T$ dependent, $s\leq 1$, and $\epsilon_0$ is the
permittivity of free space.  The dipolar contribution to the real
part of the permittivity ($\varepsilon_d^{\prime}$) has a
characteristic frequency response that is related to that of
$\sigma_d$ by the Kramers-Kronig relations,
\begin{eqnarray}
\omega\epsilon_0\varepsilon_d^{\prime}(\omega)=\sigma_d(\omega)\tan(s\pi/2).
\label{KKeqs}
\end{eqnarray}

Dipolar relaxation effects are often evidenced in
$\varepsilon_d^{\prime}$ or $\varepsilon_d^{\prime\prime}$ as
maxima at a temperature that increases with increasing $f$. These
features can be described empirically by the Cole-Cole
expression,\cite{ColeCole}
\begin{eqnarray}
\varepsilon_d =
\varepsilon_{d,\infty}+{\Delta\varepsilon_d\over 1+{(i\omega\tau)}^{1-\beta}}\ ,
\label{ColeColeRelaxation}
\end{eqnarray}
where $\varepsilon_{d,\infty}$ is the value of $\varepsilon_d$ in the high-frequency
limit, and $\Delta\varepsilon_d$ is the difference between low- and high-frequency
limiting values.  $\beta$ is an empirical parameter describing
(symmetric) relaxation broadening ($\beta=0$ corresponds to
monodispersive relaxation), and $\tau$ is the relaxation time.

\begin{figure}
\vskip -.2in
\includegraphics[width = 3.in]{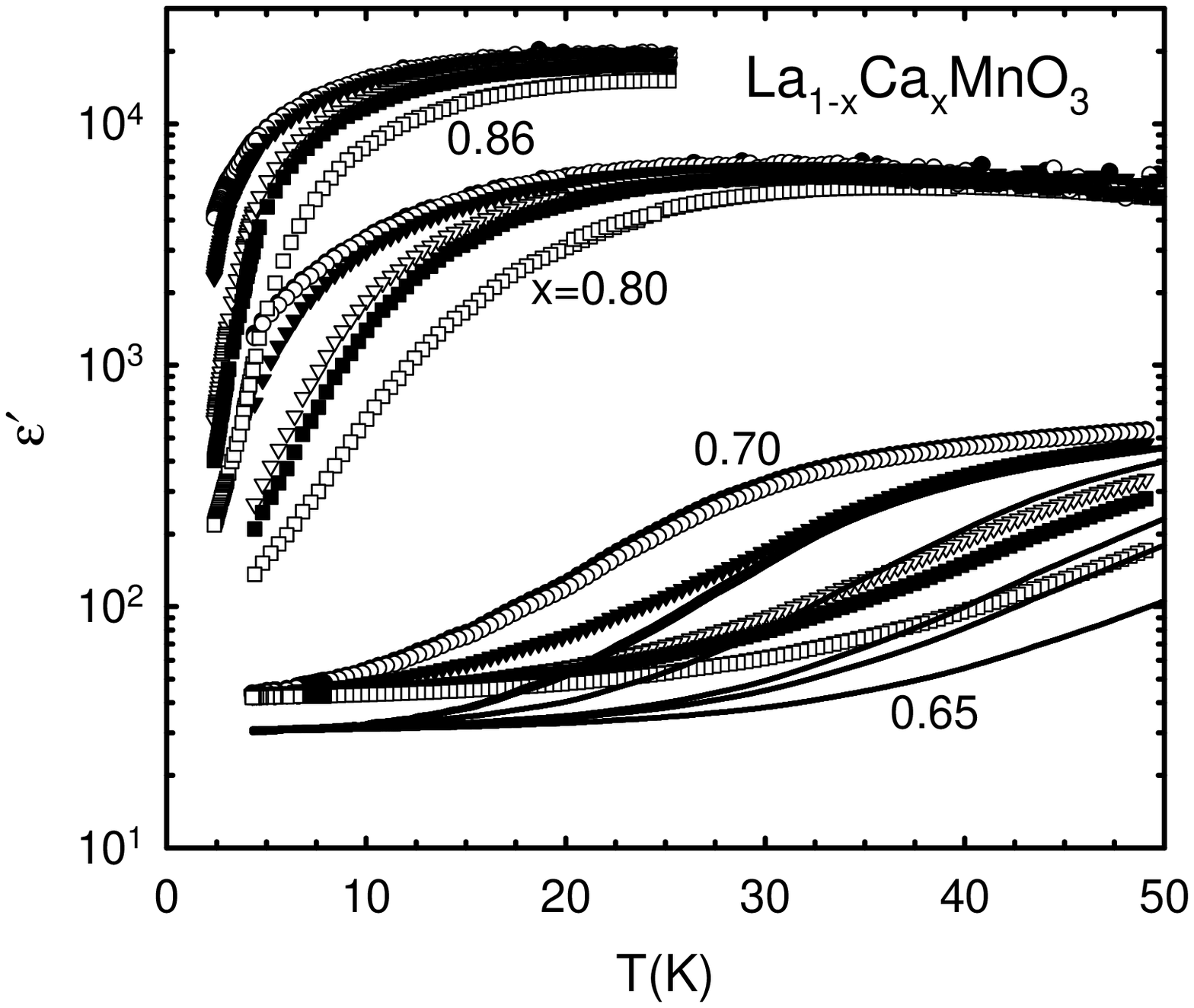}%
\vskip -2.0in
\caption{$\varepsilon^{\prime} (T)$  for LCMO
specimens.} \label{COepsdata}
\end{figure}
\begin{figure}
\vskip -.28in
\includegraphics[width = 3.75in]{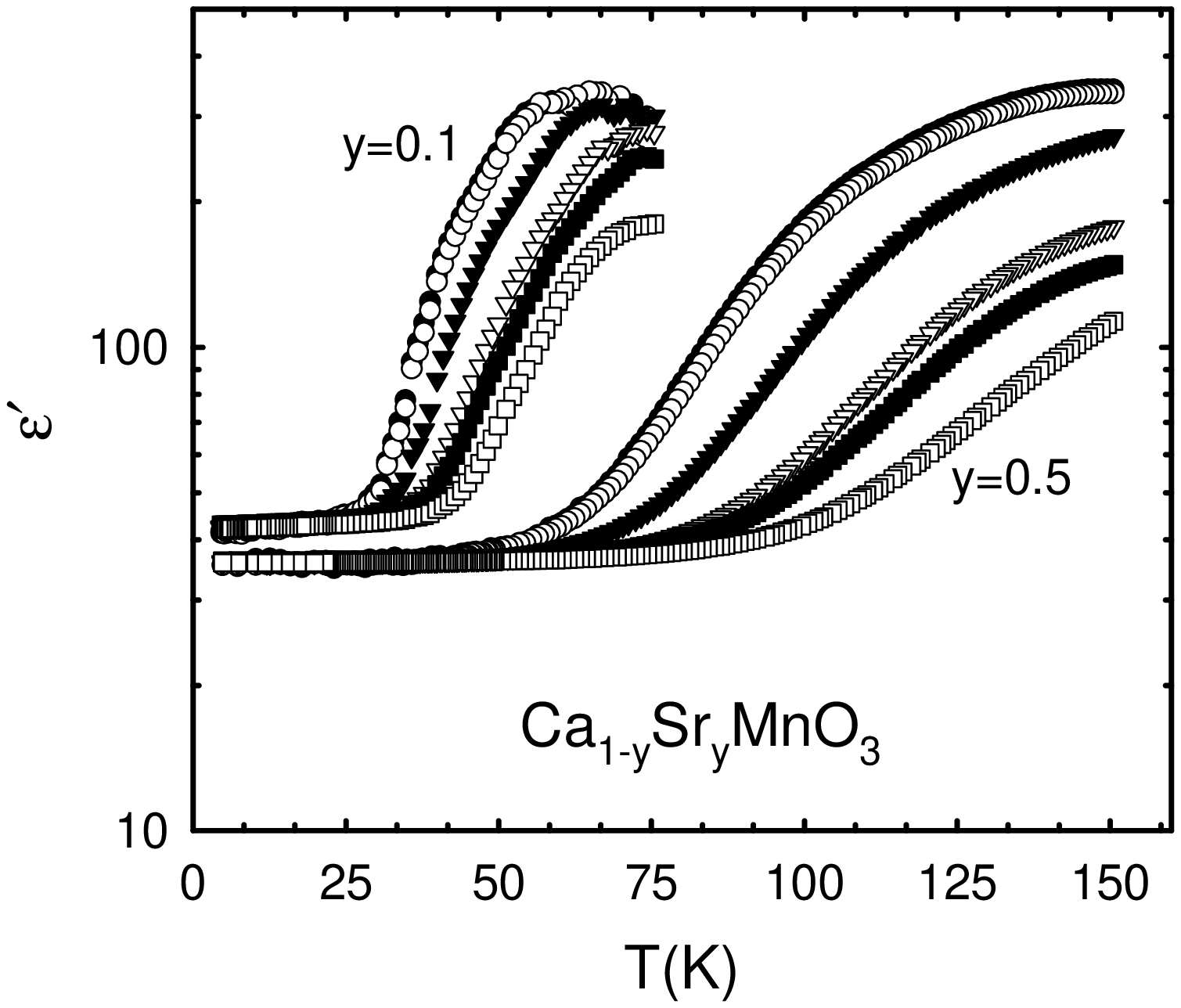}%
\vskip -2.4in
\caption{$\varepsilon^{\prime} (T)$  for CSMO specimens.}
\label{CSMOepsdata}
\end{figure}

\subsection{\label{EpsofT} Temperature Dependence of Permittivity}

$\varepsilon^{\prime}(T)$ and $\sigma(T)$ are shown for the end-member
compounds, LaMnO$_3$ (LMO) and CaMnO$_3$ (CMO) in
Fig.~\ref{LMOCMOepsdata}. $\varepsilon^{\prime}(T)$ is shown for
LCMO ($x=0.65, 0.70, 0.80$) and CSMO ($y=0.1, 0.50$) in Fig.'s~\ref{COepsdata} and
Fig.~\ref{CSMOepsdata}, respectively.

The data for most of the specimens exhibit the canonical behavior
described in the preceding section; at the lowest temperatures, $\varepsilon^{\prime}$ is
independent of temperature and frequency, reflecting an intrinsic
static dielectric constant,
$\varepsilon_0\equiv\varepsilon^{\prime}(T\to 0)$.  LMO
[Fig.~\ref{LMOCMOepsdata}~(a)] has $\varepsilon_0=18$, in good
agreement with values in the range 15$-$21 reported
previously.\cite{Lunkenheimer, Alexandrov}  In this low-temperature regime, the dc
conductivity is small, and the dispersive dipolar conductivity is
apparent [Fig.~\ref{LMOCMOepsdata}~(b)]. With increasing
temperature, dispersive maxima develop in $\varepsilon^{\prime}$,
the signature of dipolar relaxation with a relaxation time $\tau$
that decreases with increasing $T$.  Two sets of relaxation maxima
are evident in the CMO data, the one at lower $T$ evident as a
``shoulder'' in the data for the range 40$-$60~K.  The
$\varepsilon^{\prime}$ data for CMO and LCMO ($x=0.80$, $0.84$) do not
reach this $T$-independent regime for $T\geq 2$~K, so
$\varepsilon_0$ must be evaluated by extrapolation.  The data for CMO
are near saturation; $\varepsilon_0=55\pm 6$ is estimated from the average of the
$T=0$ extrapolated values of $\varepsilon^{\prime}$ for $f=10$~kHz, 20 kHz, 100 kHz.
For $x=0.80$ and $0.84$ we employ a procedure that exploits the power-law frequency behavior for the
dipolar terms as described in the next subsection.

\begin{figure}
\includegraphics[width = 3.58in]{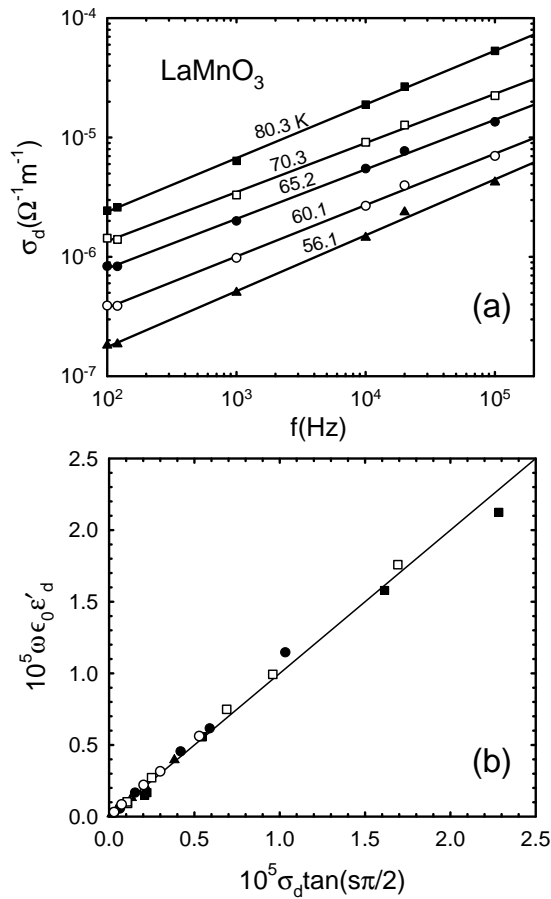}%
\vskip -.18in
\caption{(a) $\sigma_d(\omega)=\sigma(\omega)-\sigma_{dc}$ at
several fixed temperatures for LMO demonstrating power-law
behavior (Eq.~\ref{SigmaofOmega}). Solid lines are
linear-least-squares fits. (b)
$\omega\epsilon_0\varepsilon^{\prime}$ {\it vs.}
$\sigma_d\tan(s\pi/2)$ using slopes from (a). The solid line represents Eq.~\ref{KKeqs}.}
\label{KKAnalysis}
\vskip -.13in
\end{figure}

Fig.~\ref{KKAnalysis} demonstrates that the dipolar contributions to
$\varepsilon^{\prime}$ and $\sigma$ have a common origin, consistent with charge-carrier hopping.
In Fig.~\ref{KKAnalysis}~(a), linear least-squares fits
of $\sigma_d\equiv\sigma-\sigma_{dc}$ {\it vs.} $f$ in a double logarithmic plot
yield powers $s$ at various $T$'s for LaMnO$_3$.
In Fig.~\ref{KKAnalysis}~(b) these values of $s$ and
$\varepsilon_d^{\prime}\equiv\varepsilon^{\prime}-\varepsilon_0$
are used to verify Eq.~\ref{KKeqs}.
\begin{figure}
\includegraphics[width = 3.55in]{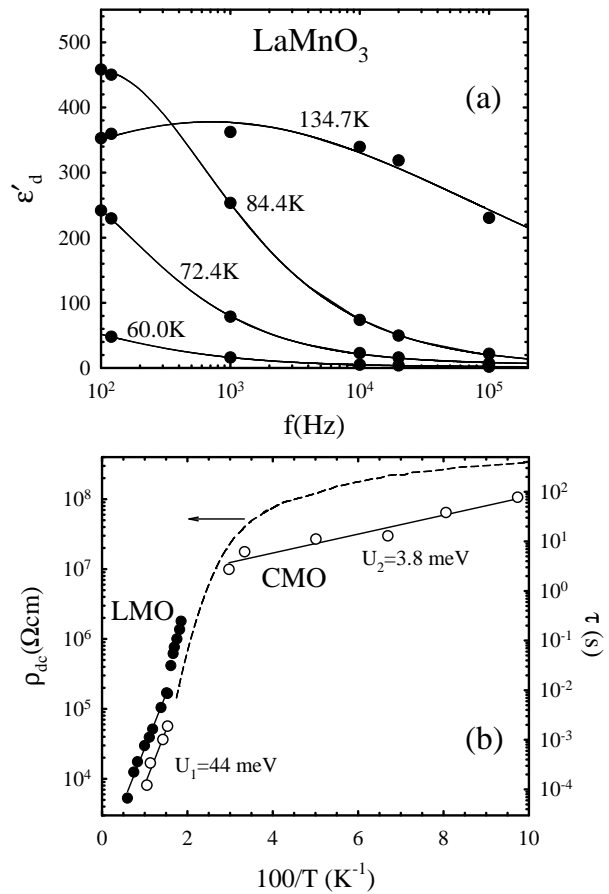}%
\vskip -.18in
\caption{(a) $\varepsilon_d^{\prime}(\omega)$ at fixed
temperatures for LMO. Curves are fits to
Eq.~\ref{ColeColeRelaxation}. (b) Fitted values of $\tau$ {\it
vs.} $100/T$ for LMO and CMO (right ordinate).  Dashed curve is
$\rho_{dc}$ for CMO (left ordinate).} \label{ColeColeLMOCMO}
\vskip -.15in
\end{figure}

Dipolar relaxation times, $\tau$, were determined for all compounds by fitting
$\varepsilon^{\prime}(\omega)$ at fixed temperatures to
Eq.~\ref{ColeColeRelaxation} as shown in
Fig.~\ref{ColeColeLMOCMO}~(a) for LMO. Values of $\beta$ fell
in the range 0.4$-$0.8, indicating a distribution of rates as is
typical for hopping systems.  $\tau$ is plotted
against inverse temperature for both LMO and CMO in
Fig.~\ref{ColeColeLMOCMO}~(b) (for CMO, data in the regime of overlap
for the two relaxation peaks were excluded).  $\tau(T)$ is approximately Arrhenius-like
in the accessible temperature ranges, $\tau=\tau_0\exp(U/k_BT)$.  For several compounds, two activation
energies, $U_1$ and $U_2$, are defined at high- and low-$T$ as observed for CMO.  The crossover
between these two relaxation regimes coincides with a crossover in the $T$ dependence of the
ac and dc resistivities [dashed curve in Fig.~\ref{ColeColeLMOCMO}~(b)].  This behavior is characteristic
of a change in the conduction mechanism from thermal activation of
carriers from impurity (dopant) levels to the conduction band at high $T$, to impurity-band conduction at low $T$.
This crossover is detectable in $\tau$ only for specimens having a sufficient carrier density to yield
a measurable dipolar contribution to $\varepsilon^{\prime}$ extending to the low-$T$ regime.
Values of activation energies and associated values of $\tau_0$ are listed in
Table~\ref{tab:Table1} for all compounds.
\begin{figure}
\includegraphics[width = 3.5in]{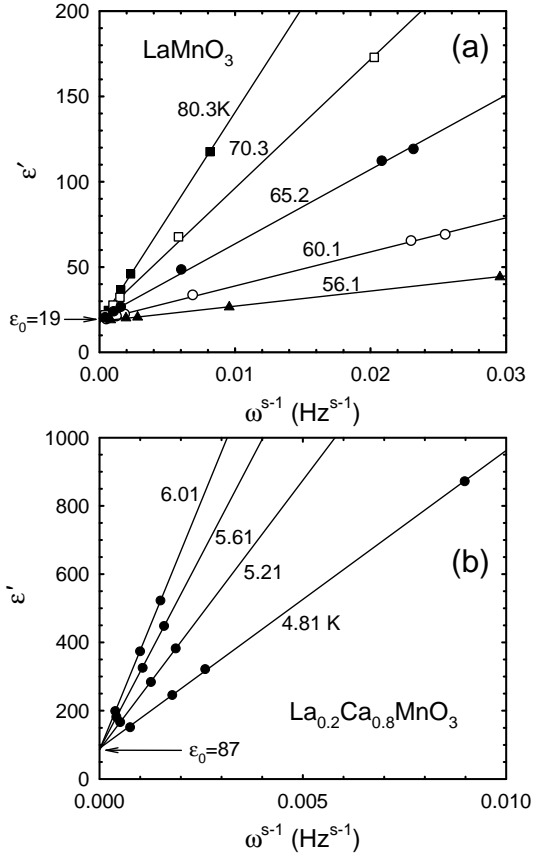}%
\caption{$\varepsilon_d^{\prime}$ {\it vs.} $\omega^{s-1}$ for (a)
LMO and (b) LCMO, $x=0.8$.} \label{EpsvsOmegaPowerLaw}
\vfill
\end{figure}
\begin{figure}
\vskip -.25in
\includegraphics[width = 3.8in]{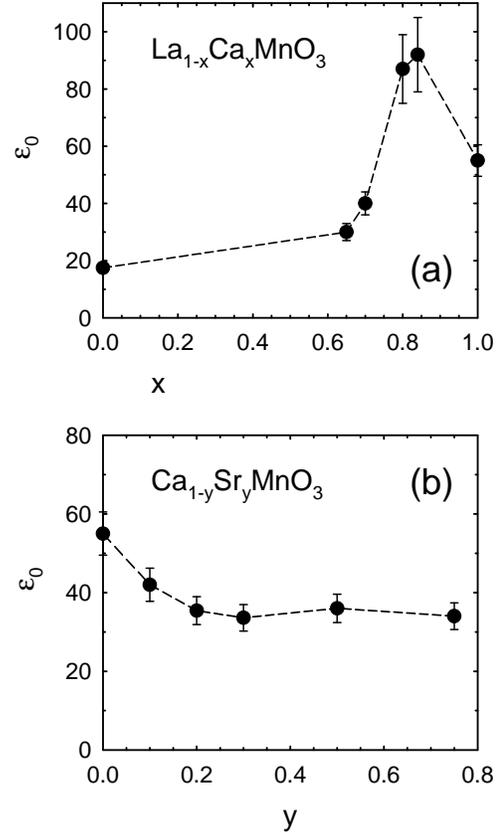}%
\vskip -.6in
\caption{Compositional dependencies of $\varepsilon_0$ for (a) LCMO  and
(b) CSMO compounds.  Error bars reflect 10\% geometric uncertainty with the exception
of LCMO $x=0.80, 0.84$ for which the extrapolation procedure (see text) yields a larger 14\%.}
\label{EpsvsXandY}
\end{figure}

\subsection{Compositional dependence of $\varepsilon_0$}

To determine $\varepsilon_0$ for LCMO $x=0.80$ and $0.84$, we employ
Eq.'s~\ref{SigmaofOmega} and \ref{KKeqs} which imply,
$\varepsilon^{\prime}=\varepsilon_0+A\omega^{s-1}$
[$A=(\sigma_0/\varepsilon_0)\tan(s\pi/2)$ is independent of
$\omega$].  Thus plots of $\varepsilon^{\prime}$ {\it vs.}
$\omega^{s-1}$ at fixed temperatures yield $\varepsilon_0$ as the common intercept (i.e., in the
limit $\omega\to\infty$).  At each temperature, $s$ is determined from the frequency dependence
of $\sigma_d$ as in Fig.~\ref{KKAnalysis}.  This procedure is validated
by application to LMO [Fig.~\ref{EpsvsOmegaPowerLaw}~(a)], using data at the same temperatures
for which $s$ was determined in Fig.~\ref{KKAnalysis}~(a).
The intercepts yield $\varepsilon_0=19\pm 2$, in good agreement with the value
$\varepsilon_0=18$ established from the low-$T$ saturation of $\varepsilon^{\prime}$
in Fig.~\ref{LMOCMOepsdata}.  Fig.~\ref{EpsvsOmegaPowerLaw}~(b)
shows results for $x=0.80$, which yield $\varepsilon_0=87\pm 12$.  A similar analysis
gives $\varepsilon_0=92\pm 13$ for $x=0.84$.  The compositional
dependencies of $\varepsilon_0$ for both the LCMO and CSMO
compounds are shown in Fig.~\ref{EpsvsXandY} and Table~\ref{tab:Table1}.
\begin{table}
\vglue -.2in
\caption{\label{tab:Table1} Static dielectric constant
$\varepsilon_0$, activation energies
describing dipolar contribution from charge-carrier hopping, $U_1$
(high-$T$) and $U_2$ (low-$T$), and corresponding values of
prefactors for Arrhenius relaxation times, $\tau_{01}$ and
$\tau_{02}$, from Cole-Cole fits of $\varepsilon^{\prime}(\omega)$
[Eq.~\ref{ColeColeRelaxation} and Fig.~\ref{ColeColeLMOCMO}].}
\begin{ruledtabular}
\begin{tabular}{cccccccc}
 &$\varepsilon_0$&$U_1$~(meV)&$U_2$~(meV)& $\tau_{01}$~(s)&$\tau_{02}$~(s)\\
\hline LaMnO$_3$&18& 44 & --- & 3.3$\times 10^{-8}$ & --- \\
La$_{0.35}$Ca$_{0.65}$MnO$_3$&31& 18 & 4.3 & 1.5$\times10^{-5}$ & 1.2 \\
La$_{0.3}$Ca$_{0.7}$MnO$_3$&40& 10& 3.9 & 8.9$\times10^{-5}$ & 5.6$\times10^{-2}$\\
La$_{0.2}$Ca$_{0.8}$MnO$_3$&87& 3.9 & 0.7 & 7.9$\times10^{-6}$
&3.3$\times10^{-4}$ \\
La$_{0.16}$Ca$_{0.84}$MnO$_3$&91& 1.3 & 0.2 & 1.1$\times10^{-4}$
&4.6$\times10^{-4}$ \\
CaMnO$_3$&55& 44 & 3.8 & 7.9$\times10^{-7}$& 1.0 \\
Ca$_{0.9}$Sr$_{0.1}$MnO$_3$&42& 33 & --- & 2.2$\times10^{-6}$ &--- \\
Ca$_{0.8}$Sr$_{0.2}$MnO$_3$&35& 33 & --- & 7.1$\times10^{-5}$  & --- \\
Ca$_{0.7}$Sr$_{0.3}$MnO$_3$&34& 31 & --- & 3.2$\times10^{-3}$& ---\\
Ca$_{0.5}$Sr$_{0.5}$MnO$_3$&36& 36 & --- & 1.7$\times10^{-3}$ & ---\\
Ca$_{0.25}$Sr$_{0.75}$MnO$_3$&34& 34 &---& 2.8$\times10^{-4}$ & --- \\
\end{tabular}
\end{ruledtabular}
\end{table}
\section{\label{sec:Discussion} Discussion}

\begin{figure}
\vskip -.1in
\includegraphics[width = 3.5in]{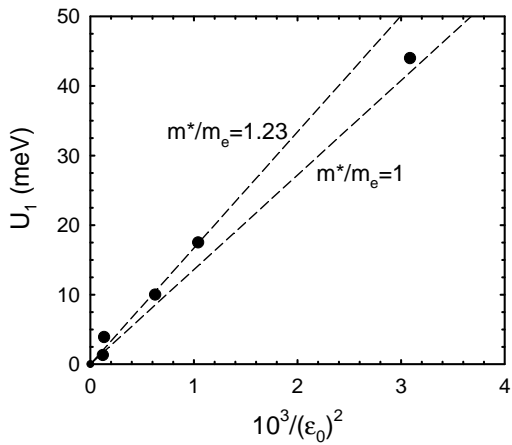}%
\vskip -2.2in
\caption{Activation energy, $U_1$ {\it vs.} inverse square of
$\varepsilon_0$ for LCMO specimens (excluding CMO). The dashed
lines represent the expectation for hydrogenic impurity levels,
with effective mass ratios indicated.} \label{HydrogenicLevels}
\end{figure}

It is evident from the data in Table~\ref{tab:Table1} that for the
LCMO compounds (excluding CMO), larger values of $\varepsilon_0$
are associated with smaller values of $U_1$.  This suggests an
interpretation within a simple model for hydrogenic impurity
levels for which the binding energy of donor (or acceptor) levels,
which we identify as $U_1$, should scale inversely with the square
of the dielectric constant,
$U_1=(m^{\ast}/m_e)(1/\varepsilon_0^2)\times 13.6$~eV.
Figure~\ref{HydrogenicLevels} demonstrates good agreement with
this simple relation for these specimens with effective mass
ratios in the range $m^{\ast}/m_e\simeq~1-1.3$.

For the nominally Mn$^{4+}$ CSMO compounds, $\varepsilon_0$ and $U_1$ are independent of composition within
uncertainties.  Using the average of these values for the five CSMO compounds in the hydrogenic impurity
expression implies $m^{\ast}/m\simeq 3.2$.  CMO, also nominally Mn$^{4+}$, appears to be an outlier.
However, there is evidence that the CMO specimen has a higher carrier density
than the CSMO compounds: both its higher low-$T$ conductivity (the CSMO compounds have conductivities similar
to that of LMO) and non-saturating $\varepsilon^{\prime}$ (Fig.~\ref{LMOCMOepsdata}).
Hall coefficient measurements\cite{CMOHall}
on a similar CMO specimen yield an electron-like Hall number at room temperature,
$n_H\simeq 2\times 10^{-4}$~f.u.$^{-1}\simeq 3\times 10^{18}$~cm$^{-3}$.
A small oxygen vacancy concentration
is a likely source of electrons, but a distribution of donors and acceptors is common in oxides.
A smaller concentration of acceptors in the present compounds is expected to arise
from several ppm levels of impurities (e.g., Al, Zn) in the starting chemicals.
Assuming this value of $n_H$ corresponds
to full ionization of impurities, we have $N_D-N_A=n_H$ where $N_D$ and $N_A$ are the donor and acceptor concentrations,
respectively.  Donors (or acceptors) with bound electrons enhance
the polarizability of a host lattice, and can plausibly account for the larger value of $\varepsilon_0$ observed
for CMO.  At low carrier density, where the donor-doped dielectric constant is not much larger than that of
the undoped host ($\varepsilon_h$), $\varepsilon-\varepsilon_h=4\pi N_D\alpha$, where $\alpha$ is the
polarizability of a single donor.  Taking $\varepsilon_h=36$
(the average value for the CSMO specimens) and $N_D=n_H$ yields the reasonable value,
$\alpha=3.8\times 10^{-19}$~cm$^3$.

Perhaps the most interesting results of the present work are the very large values of $\varepsilon_0$
observed for the two LCMO compounds, $x=0.80$ and 0.84 [Fig.~\ref{EpsvsXandY}~(a)].  Recent neutron
diffraction studies on specimens with these same compositions\cite{LingGranado}
indicate a mixture of monoclinic and orthorhombic structures at low $T$ associated with C-type AF and
Wigner-crystal (WC) type\cite{Fernandez,Radaelli} Jahn-Teller distorted, charge- and orbitally-ordered states,
respectively.  Both specimens contain approximately 80\% of the C-type phase.  The implication is
that the large values of $\varepsilon_0$ are associated with the monoclinic, C-type AF phase.
Taking the value $\varepsilon_0\simeq 31$ as representative of the WC phase
(optimized near $x=2/3$)\cite{Fernandez,Radaelli}, and assuming
measured values of $\varepsilon_0$ for $x=0.8$ and 0.84 represent weighted averages (by volume) of the
values of the two component phases, a pristine C-type polycrystal is predicted to possess an even larger,
$\varepsilon_0\sim 105$.
The increase of $\varepsilon_0$ in going from $x=0.65$ to $x=0.7$ suggests
that the $x=0.7$ specimen contains $\sim 10$~\% of the C-type monoclinic phase.
A much smaller component of the C-type phase was also detected in structural studies on a
$x=2/3$ compound.\cite{Radaelli}

In the absence of any known structural features (e.g., off-center atoms) that could enhance
$\varepsilon_0$ of the C-type phase over that of the WC phase, the substantially
lower values of $\varepsilon_0$ observed for compositions $x=0.65$ and 0.70
suggest that the one-dimensional charge/orbital ordering that characterizes the C-type phase
may play a role in determining the larger $\varepsilon_0$ found for $x=0.8$ and 0.84.
It is well-established from work on heavily doped Si\cite{DopedSiEps} and
La$_2$CuO$_{4+y}$\cite{ChenLCO} that enhancements in $\varepsilon_0$ by more than an order of magnitude
above undoped, host-lattice values are associated with the polarizability
of donors or acceptors with bound charges.  The results for the latter material may be particularly relevant
here because they demonstrate that, in a related class of AF oxides, this impurity-state polarizability enhancement
follows the electronic anisotropy.  The C-type AF state is highly anisotropic, with FM double-exchange interactions
mediating a substantially higher carrier hopping rate along the direction of $d_{3z^2-r^2}$ orbital polarization,
and superexchange
interactions suppressing hopping in the transverse directions.  It is plausible that the La-donor polarizability
is correspondingly anisotropic.  Thus the large $\varepsilon_0$ found for the $x=0.8$ and 0.84 polycrystals
could arise primarily from an enhancement of $\varepsilon_0$ along ($\|$) the FM chains of the C-type phase
[the (10$\bar{1}$) direction of the monoclinic structure], with $\varepsilon_0$ in the transverse ($\bot$)
directions comparable to that of the WC phase.  Within this scenario, the inferred $\varepsilon_0\sim 100$ for
a C-type polycrystal would represent an average, $\sim (\varepsilon_{0_{\|}}\varepsilon_{0_{\bot}})^{1/2}$, such
that $\varepsilon_{0_{\|}}\sim 300$.  To our knowledge single crystals of the C-type compositions have
not been reported, but it is clear that a study of the anisotropy of $\varepsilon_0$ in such materials would
provide further insight into the role of the orbital order in enhancing $\varepsilon_0$.

\section{\label{sec:Ack} Acknowledgments}

The work at the University of Miami was supported by NSF Grant No.
DMR-0072276 and at Montana State University by NSF Grant No. DMR-0301166.

\noindent
$^{\dagger}$ present address, Physics Department, University of Pennsylvania,
Philadelphia, PA

\end{document}